\documentstyle[12pt,twoside]{article}

\newtheorem{lem}{Lemma}

\newtheorem{cor}{Corollary}
\newtheorem{theo}{Theorem}
\newtheorem{deff}{Definition}

\def\ba{\begin{eqnarray}}
\def\ea{\end{eqnarray}}
\def\be{\begin{equation}}
\def\ee{\end{equation}}
\newfont{\msbm}{msbm10}
\newfont{\msbms}{msbm6}  
\newfont{\cmss}{cmss10}  
\def\L{{\cal L}}

\def\A{{\cal A}}
\def\S{{\cal S}}
\def\E{{\cal E}}

\def\H{{\cal H}}

\def\G{{\cal G}}

\def\hg{{\cal HG}}

\def\ag{{{\cal A}/{\cal G}}}
\def\agb{{\overline \ag}}

\def\Sig{\Sigma}

\def\Si{\mbox{$\Sigma$}}

\def\b{\beta}
\def\s{\sigma}

\def\gl{{\sl g}}

\def\h2{${\rm h}(2)$}

\def\sX{\hbox{\cmss X}}

\def\C{\hbox{\msbm C}}

\begin{document}

%%%%%%%%%%%%%%%%%%%%%%%%%%%%%%%%%%%%%%%%%%%%%%%%%%%%%%%%%%%%%%%%%%%%%%

\title{
A groupoid approach to spaces of\\generalized connections}

\author{J. M. Velhinho}

\date{{\it Centra-UAlg}\\~\\
{\footnotesize {\it Present address:} Dep. de F\'\i sica,
Universidade da Beira Interior,\\
R. Marqu\^es d'\'Avila e Bolama,
6201-001 Covilh\~a, Portugal\\
E-mail: jvelhi@mercury.ubi.pt}}

\maketitle
%%%%%%%%%%%%%%%%%%%%%%%%%%%%%%%%%%%%%%%%%%%%%%%%%%%%%%%%%%%%%%%%%5

%%%%%%%%%%%%%%%%%%%%%%%%%%%%%%%%%%%%%%%%%%%%%
\begin{abstract}

\noindent The quantum completion $\bar \A$ of the space of connections in a
manifold can be seen as the set of all morphisms from the 
groupoid of the edges of the manifold
to the (compact) gauge group. This algebraic construction  
generalizes an analogous description
of the gauge-invariant quantum configuration space $\agb$
of Ashtekar and Isham, 
clarifying the relation between the two spaces. We present a 
description of the groupoid approach
which brings the  gauge-invariant
degrees of freedom to the
foreground, thus making the action of the gauge group more transparent. 

\end{abstract}
%%%%%%%%%%%%%%%%%%%%%%%%%%%%%%%%%%%%%%%%%%%%%%
%%%%%%%%%%%%%%%%%%%%%%%%%%%%%%%%%%%%%%%%%%%%

\newpage 

%%%%%%%%%%%%%%%%%%%%%%%%%%%%%%%%%%%%%%%%%%%%
\pagestyle{myheadings}
\markboth{J.M. Velhinho}{Groupoid approach to generalized connections}
\markright{Groupoid approach to generalized connections}
%%%%%%%%%%%%%%%%%%%%%%%%%%%%%%%%%%%%%%%%%%%%%%%%%%%%%%%%%%%%%%%%%%%%%%%%%%%%

%%%%%%%%%%%%%%%%%%%%% BEGIN PAPER %%%%%%%%%%%%%%%%%%%%%%%%%

%%%%%%%%%%%%%%%%%%%%%%%%%%%%%  Section  %%%%%%%%%%%%%%%%%%%%%%%%%%%%%%%%%
\section{Introduction}
\label{int}
%%%%%%%%%%%%%%%%%%%%%%%%%%%%%%%%%%%%%%%%%%%%%%%%%%%%%%%%%%%%%%%%%%%%%%%%%

Theories of connections play an important role in the description
of fundamental interactions, including Yang-Mills theories 
\cite{Wei2}, Chern-Simons
theories \cite{Wi} and gravity in the Ashtekar formulation \cite{As}. 
Typically in such cases, the classical 
configuration space $\ag$ of connections modulo gauge transformations is
an infinite dimensional non-linear space of great complexity, challenging the
usual field quantization techniques.

Having in mind a rigorous quantization of theories of connections 
and eventually of
gravity,  methods of functional calculus in an extension
of $\ag$ were developed over the last decade. For a compact gauge group $G$,
the extension $\agb$ introduced by Ashtekar and Isham \cite{AI} is a natural
compact  measurable space, allowing the construction of well 
defined diffeomorphism
invariant measures \cite{AL1,AL2,B2}.
Like in the case of measures in infinite
dimensional linear spaces,
which appear
in the context of constructive quantum scalar
field theory, interesting
measures
in  $\agb$ are expected to be supported not
on classical configurations
but
on genuine (distributional-\- -like) generalized
connections (this was
indeed proven to be the case for the
Ashtekar-Lewandowski measure \cite{AL1},
in \cite{MM} and \cite{MTV}).

In later developments, Baez \cite{B1} considered an extension $\bar\A$ of the
space $\A$ of smooth connections. In this case one still has to
divide by the appropriate action of gauge transformations. Besides being 
equally relevant for integral calculus, the space $\bar\A$ is particularly
useful for the definition of differential calculus in $\agb$, fundamental
in the construction of quantum observables \cite{AL3}.

The construction of both $\agb$ and $\bar\A$  rely crucially on the 
use of Wilson variables
(and generalizations), bringing to the foreground the important role of
parallel transport defined by certain types of curves. In this work
we will consider only the case of piecewise analytic curves, for which the
formalism was originally introduced, although most of the arguments apply
equally well to the more general piecewise smooth case developed by Baez
and Sawin \cite{BS} and later by Lewandowski and Thiemann \cite{LT}
(see also  \cite{TW} and \cite{F1,F2} for more recent developments). 
For both $\bar\A$ and $\agb$
one considers functions on $\A$ of the form
%%%%%%%%%%%
\be
\label{seccao10.1eqB}
\A\ni A\mapsto F\bigl(h(c_1,A),\ldots,h(c_n,A)\bigr)\,,
\ee
%%%%%%%%%%%
where $h(c,A)$ denotes the parallel transport defined by the curve $c$ and
$F:G^n\to\C$ is a continuous function. In the case of $\agb$ only
closed curves -- loops -- are needed, producing
gauge invariant functions, or functions on $\ag$. These functions are
sufficient to define (overcomplete) coordinates on  $\ag$ \cite{AI}.
For compact $G$, the set of all functions (\ref{seccao10.1eqB})
is naturally a normed commutative $\ast$-algebra with identity. The 
completion of such an algebra is therefore a commutative  unital 
$C^{\ast}$-algebra and, according to Gelfand theory, this $C^{\ast}$-algebra
can be seen as the algebra of continuous functions on a compact
space called the spectrum of the algebra. The spectrum of the above
algebras -- $\agb$ for the closed curves case and $\bar\A$ for the general open
curves case -- are natural completions of $\ag$ and $\A$, respectively, 
and appear as
good candidates to replace them in the quantum context.

To a large extent, the definition of functional calculus on $\agb$ rely
on the fact that, while being extremely complex spaces,
both $\agb$ and $\bar\A$ 
can be seen as projective limits
of families of finite dimensional compact manifolds \cite{AL1,MM,AL2}
(see also \cite{B1,BS,LT} for a formulation in terms of inductive
limits). This projective characterization gives us a great deal of control
over the spaces $\agb$ and $\bar\A$, 
allowing the construction of
measures and vector fields starting from  corresponding
structures on the compact finite dimensional spaces in the projective 
families \cite{AL1,AL2,MM,B1,AL3}.

The projective approach leads also to an interesting interpretation of
generalized connections. For the case of $\agb$, a distinguished group 
of equivalence classes of loops,
called the hoop group $\hg$ \cite{AL1}, plays an important role, in the sense
that  $\agb$ can be identified with the space $\rm{Hom}\, [\hg,G]/G$ 
of all homomorphisms (modulo conjugation) from $\hg$ to $G$,
with the topology on $\rm{Hom}\, [\hg,G]/G$ being induced by a projective
family labeled by finitely generated subgroups of $\hg$.
As pointed out by Baez \cite{B3}, for $\bar\A$ a similar
role is played by a certain groupoid. In our opinion however, 
this groupoid associated
to open curves has not yet occupied the place it deserves in the literature,
possibly due to the fact that groupoids have been introduced in the current 
mathematical physics literature only recently. Recall that a
groupoid is a category such that all arrows are invertible. 
Therefore, a groupoid
generalizes the notion of a group, in the sense that a binary operation with
inverse is defined, the difference being that not all pairs of elements can\
be composed.

In section \ref{sec2} of this work we consider the 
projective characterization of $\bar\A$ 
using the language of groupoids  from the very beginning.
This amounts to putting the usual approach using graphs \cite{AL3}
in an appropriate algebraic framework, in a natural generalization of
the hoop group approach. Using this
formalism, we show in section \ref{sec3} that the quotient of $\bar\A$ by 
the action of the gauge group is 
homeomorphic to $\agb$. This new proof, establishing directly the equivalence 
at the projective limit level, seems to us more transparent than the proof 
one can obtain by combining results from \cite{AL1,MM,AL2,B1,AL3}.

%%%%%%%%%%%%%%%%%%%%%%%%%%%%%  Section %%%%%%%%%%%%%%%%%%%%%%%%%%%%
\section{Groupoid-projective formulation of $\bar \A$}
\label{sec2}
\subsection{Edge groupoid}
\label{seg}
%%%%%%%%%%%%%%%%%%%%%%%%%%%%%%%%%%%%%%%%%%%%%%%%%%%%%%%%%%%%%%%%%%%%%%%%%
Let \Si\ be an analytic, connected and orientable $d$-manifold. Let us
consider the set $\E$ of all continuous, oriented and piecewise analytic
parametrized curves in \Si , i.e.~maps
%%%%%%%%%%%%%
$$
c:[0,t_1]\cup\ldots\cup [t_{n-1},1]\to\Sig
$$
%%%%%%%%%%%%
which are continuous in all the domain $[0,1]$, analytic in the closed
intervals $[t_k,t_{k+1}]$ and such that the images $c\bigl(]t_k,t_{k+1}[
\bigr)$ of the open intervals $]t_k,t_{k+1}[$ are submanifolds
embedded in \Si . In the set $\E$ of all such curves one may define the
following maps. Let $\s:\E\to\Si$ be the map given by $\s(c)=c\bigl([0,1]
\bigr)$, $c\in\E$. The maps $s$ (source) and $r$ (range) are
defined, respectively, by $s(c)=c(0)$, $r(c)=c(1)$. Given two curves
$c_1,c_2\in\E$ such that $s(c_2)=r(c_1)$, let $c_2c_1\in\E$
denote the natural composition given by
%%%%%%%%%%%
$$
(c_2c_1)(t)=\left\{\begin{array}{lll} c_1(2t), & {\rm for} & t\in[0,1/2] \\
c_2(2t-1), & {\rm for} & t\in[1/2,1]\,. \end{array} \right.
$$
%%%%%%%%%%%
This composition defines a binary operation in a well defined subset of
$\E\times\E$. Consider also the operation $c\mapsto c^{-1}$ 
given by $c^{-1}(t)=c(1-t)$. Strictly speaking, the composition
of parametrized curves is not associative, since the curves $(c_3c_2)c_1$ 
and $c_3(c_2c_1)$ are related by a reparametrization, i.e.~by an
orientation preserving piecewise analytic diffeomorphism $[0,1]\to [0,1]$.
Similarly, the curve $c^{-1}$ is not the inverse of the curve $c$.
Following Isham, Ashtekar and Lewandowski \cite{AI,AL1} and Baez \cite{B3},
we describe next an appropriate equivalence relation in $\E$. 
The corresponding
set of equivalence classes is a well defined groupoid \cite{B3}, generalizing 
the group of hoops introduced by Ashtekar and Lewandowski \cite{AL1}.

Let $G$ be a (finite dimensional) connected and compact Lie group and let
$P(\Sig ,G)$ be a principal $G$-bundle over \Si . For simplicity we assume
that the bundle is trivial and that a fixed trivialization has been chosen.
Let $\A$ be the space of smooth connections on this bundle.
The parallel transport associated with a given connection $A\in\A$ and
a given curve $c\in\E$ will be denoted by $h(c,A)$.
%%%%%%%%%%%%%%%%%
\begin{deff}
\label{def1}
Two elements $c$ and $c'$ of $\E$ are said to be equivalent if
\begin{itemize}
\item[{\rm (}i\/{\rm )}] $s(c)=s(c')\, ,\ r(c)=r(c')\, ;$
\item[{\rm (}ii\/{\rm )}] $h(c,A)=h(c',A)$, $\forall A\in\A$.
\end{itemize}
\end{deff}
%%%%%%%%%%%%%%%%%%
It is obvious that two curves related by a reparametrization are equivalent.
Two curves $c$ and $c'$ which can be written in the form $c=c_2c_1$,
$c'=c_2c_3^{-1}c_3c_1$ are also equivalent. It can be shown that, for
compact noncommutative Lie groups G, these two conditions are equivalent to
({\it ii\/}) (see e.g.~\cite{AL2,LT}). Thus, in the context
of noncommutative compact Lie groups, the equivalence relation above 
is independent of the group.

We will consider noncommutative groups from now on and denote the set of
all above defined equivalence classes by $\E\G$.
It is clear by ({\it i\/}) that the maps $s$ and $r$ are well
defined in $\E\G$. 
The map $\s$ can still be defined for special elements called edges. 
By edges we mean elements $e\in\E\G$ which are equivalence classes of
analytic (in all domain) curves $c:[0,1]\to\Si$. It is clear that the images
$c_1\bigl([0,1]\bigr)$ and $c_2\bigl([0,1]\bigr)$ of two equivalent analytic
curves coincide, and therefore we define $\s(e)$ as being $\s(c)$, where
$c$ is any analytic curve in the classe of the edge $e$.

We discuss next the natural groupoid structure on the set $\E\G$.
We will follow the terminology of category theory and refer to elements
of $\E\G$ as arrows.

The composition of arrows
is defined by the composition of elements of $\E$:  if $\gamma,\gamma'\in\E\G$ 
are such that~$r(\gamma)=s(\gamma')$ one defines $\gamma'\gamma$ as the 
equivalence class of $c'c$, where $c$ (resp.~$c'$) belongs to the class
$\gamma$ ($\gamma'$). The independence of this composition with
respect to the
choice of representatives follows from $h(c'c,A)=h(c',A)h(c,A)$ and
from condition ({\it ii\/}) above. The composition 
in $\E\G$ is now associative,
since $(c_3c_2)c_1$ and $c_3(c_2c_1)$ belong to the same equivalence class.

The points of $\Si$ are called objects
in this context.
Objects are in one-to-one correspondence with identity
arrows:  given $x\in\Sig$ the corresponding identity ${\bf 1}_x\in\E\G$ is the
equivalence class of $c^{-1}c$, with $c\in\E$ such that $s(c)=x$. 
If $\gamma$ is 
the class of $c$ then $\gamma^{-1}$ is the class of $c^{-1}$. It is clear that
$\gamma^{-1}\gamma={\bf 1}_{s(\gamma)}$ and 
$\gamma\gamma^{-1}={\bf 1}_{r(\gamma)}$. 

One therefore has a well defined 
groupoid, whose set of objects is $\Si$ and whose set of arrows is $\E\G$.
As usual, we will use the same notation -- $\E\G$ -- both for
the set of arrows and for the groupoid.
Notice that
every element $\gamma\in\E\G$ can be obtained as a composition of edges.
Therefore, the groupoid $\E\G$ is generated by the set of edges, although
it is not freely generated, since composition of edges may produce new edges.

For  $x,y\in\Si$, let
%%%%%%%%%%%%%%%%%%%%%%%%%%%%%%%%%%%%%%%%%%%%%%%%%
\be
\label{grupg2}
{\rm Hom}\,[x,y]:=\bigl\{\gamma\in\E\G\ |\ \ s(\gamma)=
x,\ r(\gamma)=y\bigr\}
\ee
%%%%%%%%%%%%%%%%%%%%%%%%%%%%%%%%%%%%%%%%%%%%%%%%%%%%%
be the set of all arrows starting at $x$ and ending at $y$.
It is clear that ${\rm Hom}\,[x,x]$ is a group, $\forall x\in\Si$. 
Since the manifold \Si\ is taken
to be connected, the groupoid $\E\G$ is also connected, i.e.~${\rm Hom}\,
[x,y]$ is a non-empty set, for every pair $x,y\in\Sig$. In this case,
any two groups ${\rm Hom}\,[x,x]$ and ${\rm Hom}\,[y,y]$ are isomorphic.
Let us fix a point $x_0\in\Sig$ and consider the group ${\rm Hom}\,[x_0,x_0]$.
This group is precisely the so-called hoop group $\H\G$~\cite{AL1}, 
whose elements are equivalence classes 
of piecewise analytic loops. 
The elements of
${\rm Hom}\,[x_0,x_0]$ are called hoops and the identity 
arrow ${\bf 1}_{x_0}$ will be called the trivial hoop.

Given that $\E\G$ is connected, its elements may be written as
compositions of elements of ${\rm Hom}\,[x_0,x_0]$ and of an appropriate
subset of the set of all arrows:
%%%%%%%%%%%%%%%%%%%%%%
\begin{lem}
\label{lem1}
Suppose that an unique arrow $\gamma_x\in{\rm Hom}\,[x_0,x]$ is given
for each $x\in\Sig$, $\gamma_{x_0}$ being the trivial hoop.
Then for every $\gamma\in\E\G$ there is a uniquely defined 
$\b\in{\rm Hom}\,[x_0,x_0]$ such that
%%%%%%%%%%%%%
\be
\label{grupg4a}
\gamma=\gamma_{r(\gamma)}\b\gamma_{s(\gamma)}^{-1}\,.
\ee
%%%%%%%%%%%%%
%%%%%%%%%%%%%%%%%
\end{lem}
%%%%%%%%%%%%%%%%%%%%%%
This result can  be obviously adapted for any connected subgroupoid 
$\Gamma\subset\E\G$. The
converse of this result is the following lemma, where
${\rm Hom}_{\,\Gamma}\,[x_0,x_0]$ denotes the subgroup of the hoops that belong
to $\Gamma$.
%%%%%%%%%%%%%%%%%%%%%%%%%
\begin{lem}
\label{lem2}
Let $F$ be a subgroup of ${\rm Hom}\,[x_0,x_0]$ and $X\subset\Sig$ be a
subset of \Si\ such that~$x_0\in X$. Suppose that an unique arrow
$\gamma_x\in{\rm Hom}\,[x_0,x]$ is given for each $x\in X$, 
$\gamma_{x_0}$ being the trivial hoop. Then the set $\Gamma$ of 
all arrows of the form
$\gamma_x\b\gamma_y^{-1}$, with $\b\in F$ and 
$x,y\in X$, is a connected subgroupoid of $\E\G$, and the group
${\rm Hom}_{\,\Gamma}\,[x_0,x_0]$ coincides with $F$.
\end{lem}
%%%%%%%%%%%%%%%%%%%%%%%%%%%
To prove that $\Gamma$ is subgroupoid it is sufficient to show that
{\it i\/}) every arrow $\gamma\in\Gamma$ is invertible in $\Gamma$ and
{\it ii\/}) that the composition $\gamma\gamma'$ belongs to $\Gamma$, for
every $\gamma,\gamma'\in\Gamma$ such that~$\gamma\gamma'$ is defined on
$\E\G$. The inverse of $\gamma_x\b
\gamma_y^{-1}$ is $\gamma_y\b^{-1}\gamma_x^{-1}
\in\Gamma$, proving {\it i\/}). As for {\it ii\/}), notice that given
$\gamma=\gamma_x\b\gamma_y^{-1}$ and
$\gamma'=\gamma_{x'}\b'\gamma_{y'}^{-1}$, the composition
$\gamma\gamma'$ is defined if and only if $y=x'$, and therefore 
$\gamma\gamma'=\gamma_x(\b\b')\gamma_{y'}^{-1}$ 
belongs to $\Gamma$,
since  $F$ is a group. The groupoid $\Gamma$ is connected, given that 
every object $x\in X$ is connected to $x_0$ by an arrow. If $\gamma=
\gamma_x\b\gamma_y^{-1}$ belongs to ${\rm Hom}_{\,\Gamma}\,
[x_0,x_0]$ then $x=y=x_0$ and $\gamma=\b\in F$. Conversely,
it is obvious that $F\subset{\rm Hom}_{\,\Gamma}\,[x_0,x_0]$.~$\Box$

%%%%%%%%%%%%%%%%%%%%%%%%%%% Section %%%%%%%%%%%%%%%%%%%%%%%%%%%%%%%%%%%%%%%%
\subsection{$\bar \A$ as a projective limit}
\label{ssi}
%%%%%%%%%%%%%%%%%%%%%%%%%%%%%%%%%%%%%%%%%%%%%%%%%%%%%%%%%%%%%%%%%%%%%%%%%%%%

By the very definition of $\E\G$ (see condition ({\it ii\/}) in 
definition \ref{def1}),
the parallel transport is well defined for any element of $\E\G$. To
emphasize the algebraic role of connections and to simplify the notation,
we will denote by $A(\gamma)$ the parallel transport $h(c,A)$ defined
by $A\in\A$ and any curve $c$ in the equivalence class $\gamma\in\E\G$. Let us
recall that the bundle $P(\Sig,G)$ is assumed to be trivial, and
therefore $A(\gamma)\equiv h(c,A)$ defines an element of the group $G$.
For every connection $A\in\A$, the map from $\E\G$ to $G$ given by
%%%%%%%%%%%%%%%%%%%%%%%%%%%%%%%%%%%%%%%%%%%%%%%%
\be
\label{grupg4}
\gamma \mapsto A(\gamma)
\ee
%%%%%%%%%%%%%%%%%%%%%%%%%%%%%%%%%%%%%%%%%%%%%%%
is a groupoid morphism, i.e., $A(\gamma'\gamma)=A(\gamma')A(\gamma)$ and 
$A(\gamma ^{-1})=A(\gamma)^{-1}$. Thus, there is a well defined injective but 
not surjective \cite{AI,AL1,B1,Le1} map from $\A$ to the set ${\rm Hom}\,
[\E\G,G]$ of all morphisms from $\E\G$ to $G$, through which $\A$ can
be seen as a proper subset of ${\rm Hom}\,[\E\G,G]$. 
It turns out that ${\rm Hom}\,[\E\G,G]$, 
when equipped with an appropriate topology,
is homeomorphic to the space $\bar\A$ of generalized 
connections \cite{MM,AL2,B3}.
This identification can be proved using the fact that 
${\rm Hom}\,[\E\G,G]$ is the projective limit of a projective family
labeled by graphs in the 
manifold $\Si$ \cite{ALMMT,AL3}.
In what follows we will rephrase the projective characterization of 
${\rm Hom}\,[\E\G,G]$
using the language of groupoids. 
We start with the  set of labels for the projective family leading to
${\rm Hom}\,[\E\G,G]$, using
the notion of  independent edges \cite{AL1}.
%%%%%%%%%%%%%%%%%%%%%%%%%%%
\begin{deff}
\label{def2}
A finite set $\{e_1,\ldots,e_n\}$ of edges is said to be independent if the
edges $e_i$ can intersect each other only at the points $s(e_i)$ or 
$r(e_i)$, $i=1,\ldots,n$.
\end{deff}
%%%%%%%%%%%%%%%%%%%%%%%%%%%%
The edges in an independent set are, in particular, algebraically independent,
i.e.~it is not possible to produce identity arrows by (nontrivial) 
compositions of the edges and their inverses. Our condition of independent 
sets is of course stronger than the condition of algebraic independence.

Let us denote by $\E\G\{e_1,\ldots,e_n\}$ the subgroupoid of $\E\G$ 
generated by the independent set $\{e_1,\ldots,e_n\}$, 
i.e.~$\E\G\{e_1,\ldots,e_n\}$ is the smallest 
subgroupoid containing all the 
edges $e_i$, or explicitly, the subgroupoid whose objects are all the
points $s(e_i)$ and $r(e_i)$ and whose arrows are all possible
compositions of edges  $e_i$ and their inverses. Groupoids of this
type are freely generated, given the algebraic independence of the edges.

In what follows we will denote by $\L$ the set of all subgroupoids for
which there exists a finite set of independent edges 
such that $L=\E\G\{e_1,\ldots,e_n\}$. Clearly, the sets 
$\{e_1,\ldots,e_n\}$ and
$\bigl\{e_1^{\epsilon_1},\ldots,e_n^{\epsilon_n}\bigr\}$, where
$\epsilon_j=\pm 1$ (i.e.~$e_j^{\epsilon_j}=e_j$ or $e_j^{-1}$) generate
the same subgroupoid, and this is the only ambiguity in the choice of the
set of generators of a given groupoid $L\in\L$. Thus, a groupoid
$L\in\L$ is uniquely defined by a set $\bigl\{\s(e_1),\ldots,\s(e_n)\bigr\}$
of images of a set of independent edges. Notice that the union of the images 
$\s(e_i)$ is a graph in the manifold \Si , thus establishing the relation
with the approach used in \cite{B1,B2} and \cite{AL3}.

Let us consider in the set $\L$ the partial order relation defined by 
inclusion, i.e.~given $L,L'\in\L$, we will say that $L'\geq L$ if and only
if $L$ is a subgroupoid of $L'$. Recall that $L$ is said to be a subgroupoid 
of $L'$ if and only if all objects of $L$ are objects of $L'$ and for
any pair of objects $x,y$ of $L$ every arrow from $x$ to $y$ is an arrow
of $L'$. It is easy to see that $\L$ is a
directed set with respect to the latter partial order, meaning that 
for any given
$L$ and $L'$ in $\L$ there exists $L''\in\L$ such that
$L''\geq L$ and $L''\geq L'$.
We will not repeat here the arguments leading to this conclusion; 
the crucial fact is that for
every finitely generated subgroupoid $\Gamma\subset\E\G$ there is an
element $L\in\L$ such that~$\Gamma$ is a subgroupoid of $L$,
which can be easily proved in the piecewise analytic case \cite{AL1}.

Let us now consider the projective family. For each $L\in \L$,
let $\A_L:={\rm Hom}\,[L,G]$ be the set of all morphisms from the groupoid
$L$ to the group $G$. We will show next that the family of spaces
$\A_L,\ L\in\L$, is a so-called compact Hausdorff projective family
(see \cite{AL2}), meaning that each of the spaces $\A_L$ is a compact
Hausdorff space and that given $L,L'\in\L$ such that~$L'\geq L$ there exists a
surjective and continuous projection $p_{L,L'}:\A_{L'}\rightarrow \A_L$
such that
%%%%%%%%%%%%%%%%%%%%%%%%%%%%%%%%%%%%%%%%
\be
\label{grupg6}
p_{L,L''}=p_{L,L'}\circ p_{L',L''},\ \forall L''\geq L'\geq L\,.
\ee
%%%%%%%%%%%%%%%%%%%%%%%%%%%%%%%%%%%%%%
There is a well defined notion of limit of the family of spaces
$\A_L$ -- the projective limit -- which is also a compact Hausdorff
space.

Given $L\in\L$, let $\{e_1,\ldots,e_n\}$ be a set of independent edges that 
freely generates the groupoid $L$. Since the morphisms $L\to G$ are
uniquely determined by the images of the generators of $L$, one gets a
bijection $\rho_{e_1,\ldots,e_n}:\A_L\to G^n$, given by
%%%%%%%%%%%%
\be
\label{bara4}
\A_L\ni\bar A\mapsto \bigl(\bar A(e_1),\ldots,\bar A(e_n)\bigr)\in G^n\,.
\ee
%%%%%%%%%%%
Through this identification with $G^n$, the space $\A_L$ acquires a topology
with respect to which it is a compact Hausdorff space. Notice that the 
topology induced
in $\A_L$ is independent of the choice of the generators (including
ordering), since maps of the form
%%%%%%%%%%%%
\be
\label{bara5}
(g_1,\ldots,g_n)\mapsto \Bigl(g_{k_1}^{\epsilon_{k_1}},\ldots,
g_{k_n}^{\epsilon_{k_n}}\Bigr)\,,
\ee
%%%%%%%%%%%
where $(k_1,\ldots,k_n)$ is a permutation of $(1,\ldots,n)$ and
$\epsilon_{k_i}=\pm 1$, are homeomorphisms $G^n\to G^n$. For 
$L'\geq L$ let us define the projection $p_{L,L'}:\A_{L'}\to \A_L$ as the map
that sends each element of $\A_{L'}$ to its restriction to $L$. It is clear
that (\ref{grupg6}) is satisfied. We will now show that the maps $p_{L,L'}$ 
are surjective and continuous. Let $\{e_1,\ldots,e_n\}$ 
be generators of $L$ and $\{e'_1,\ldots,e'_m\}$ 
be generators of $L'\geq L$. Let us consider the decomposition of the edges
$e_i$ in terms of the edges $e'_j$:
%%%%%%%%%%%
\be
\label{bara1}
e_i=\prod_j ({e_{r_{ij}}'})^{\epsilon_{ij}},\ \ i=1,\ldots,n\, ,
\ee
%%%%%%%%%%%%
where $r_{ij}$ and $\epsilon_{ij}$ take values in the sets
$\{1,\ldots,m\}$ and $\{1,-1\}$, respectivelly. An arbitrary element 
of $\A_L$ is
identified by the images $(h_1,\ldots,h_n)\in G^n$ of the ordered set
of generators $(e_1,\ldots,e_n)$. The map $p_{L,L'}$ will be surjective
if and only if there are $(g_1,\ldots,g_m)\in G^m$ such that
%%%%%%%%%%%%%
\be
\label{bara2}
h_i=\prod_jg^{\epsilon_{ij}}_{r_{ij}},\ \ \forall i\,.
\ee
%%%%%%%%%%%%%
These conditions can indeed be satisfied, since they are independent.
In fact, since the edges $\{e_1,\ldots,e_n\}$ are independent, a given edge
$e'_k$ can appear at most once (in the form  $e'_k$ or $e'^{-1}_k$)
in the decomposition (\ref{bara1}) of a given $e_i$. To prove continuity
notice that, through the identification (\ref{bara4}), the map $p_{L,L'}$ 
corresponds to the projection $\pi_{n,m}:G^m\to G^n$:
%%%%%%%%%%%%%
\be
\label{bara3}
G^m\ni(g_1,\ldots,g_m)\stackrel{\pi_{n,m}}{\longmapsto}\Bigl({\textstyle
\prod_j}g^{\epsilon_{1j}}_{r_{1j}},\ldots,{\textstyle \prod _j}g^
{\epsilon_{nj}}_{r_{nj}}\Bigr)\in G^n\,,
\ee
%%%%%%%%%%%%
which is continuous.

The projective limit of the family $\{\A_L,p_{L,L'}\}_{L,L'\in\L}$ is the
subset $\A_{\infty}$ of the cartesian product \sX$_{L\in\L}\A_L$ of those
elements $(A_L)_{L\in\L}$ satisfying the following consistency conditions: 
%%%%%%%%%%%%%%
\be
\label{eq1}
p_{L,L'}A_{L'}=A_L\ ,\ \ \ \forall \ L'\geq L\,.
\ee
%%%%%%%%%%%%%%%
The cartesian product is a compact Hausdorff space with the Tychonov product
topology. Given the continuity of the projections $p_{L,L'}$, the projective 
limit $\A_{\infty}$ is a closed subset \cite{MM,AL2} and therefore is also
a compact Hausdorff space. Explicitly, the induced topology in $\A_{\infty}$
is the weakest topology such that all the following  projections
are continuous:
%%%%%%%%%%%%%%%
\ba
p_L:\qquad\quad \A_{\infty} & \to & \A_L \nonumber \\
\label{eqa1}
(A_L)_{L\in\L} & \mapsto & A_L\, .
\ea
%%%%%%%%%%%%%%%
The proof that the projective limit $\A_{\infty}$ coincides with the set of 
all groupoid morphisms ${\rm Hom}\,[\E\G,G]$ follows essentially the same
steps as  the proof of the well known fact that the algebraic dual
of any vector space is a projective limit, and therefore will not be presented
here (see e.g.~\cite{AL1,AL2,MM} for the closely related case of the space
of generalized connections modulo gauge transformations). 
It is interesting to note that ${\rm Hom}\,[\E\G,G]$ can be seen as being
dual (in a non-linear sense) to the groupoid $\E\G$. In what follows we will 
identify $\A_{\infty}$ with ${\rm Hom}\,[\E\G,G]$. For simplicity,
we will refer to the induced topology on ${\rm Hom}\,[\E\G,G]$ as the
Tychonov topology.
%%%%%%%%%%%%%%%%%%%%%%%%%%% Section %%%%%%%%%%%%%%%%%%%%%%%%%%%%%%%%%%%%%%%%
\section{Relation between $\bar\A$ and $\agb$ in the groupoid-projective 
approach}
\label{sec3}
%%%%%%%%%%%%%%%%%%%%%%%%%%%%%%%%%%%%%%%%%%%%%%%%%%%%%%%%%%%%%%%%%%%%%%%%%%%%
In this section we will study the relation between the space of
generalized connections considered above and the space $\agb$ of
generalized connections modulo gauge transformations \cite{AL1,AL2},
{}from the point of view of projective techniques. The gauge
transformations act naturally in ${\rm Hom}\,[\E\G,G]$ and, as expected,
the quotient of ${\rm Hom}\,[\E\G,G]$ by this action is homeomorphic to
$\agb$. The proof presented here complements the results in
\cite{AL1,AL2,MM,B1,AL3} and clarifies the relation between the
two spaces. The introduction of the groupoid $\E\G$ plays a relevant 
simplifying role in this result.
%%%%%%%%%%%%%%%%%%%%%%%%%%%%%%%%%%%%%%%%%%%%%%%%%%%%%%%%%%%%%%%%%%%%%%%
\subsection{Gauge transformations, $\bar\A$ and $\agb$}
\label{sgt}
%%%%%%%%%%%%%%%%%%%%%%%%%%%%%%%%%%%%%%%%%%%%%%%%%%%%%%%%%%%%%%%%%%%%%%%
We start with a brief review of the projective characterization of $\agb$
\cite{AL1,AL2,MM}. A finite set of hoops $\{\b_1,\ldots,\b_n\}$ is said
to be independent if each hoop $\b_i$ contains an edge which is traversed
only once and which is shared by any other hoop at most at a finite
number of points. In the hoop formulation the projective family
is labeled by certain ``tame'' subgroups of the hoop group $\H\G\equiv
{\rm Hom}\,[x_0,x_0]$, which are subgroups freely generated by 
finite sets of independent hoops. We will 
denote the family of such subgroups by $\S_{\H}$. For each $S\in\S_{\H}$
one considers the set $\chi_S$ of all homomorphisms $S\to G$
%%%%%%%%%%%%%%%%%%%%%%%%%%%%%%%%%
\be
\label{hoopint1}
\chi_S:={\rm Hom}[S,G]\,.
\ee
%%%%%%%%%%%%%%%%%%%%%%%%%%%%%%%%%%%%
The sets $\chi_S$ can be
identified with powers of $G$ and the family $\{\chi_S\}_{S\in\S_{\H}}$ is a
compact Hausdorff projective family, whose projective limit is 
${\rm Hom}\,[\H\G,G]$,
the set of all homomorphisms from the $\hg$ to $G$ \cite{AL2}.
By means of the projective family, the space ${\rm Hom}\,[\H\G,G]$
is equiped with a Tychonov-like topology, namely the
weakest topology such that all the natural projections
%%%%%%%%%%%%
\be
\label{hoopint4}
p_S:{\rm Hom}\,[\H\G,G]\to\chi_S\,,\ \ S\in\S_{\H}\,,
\ee
%%%%%%%%%%%%%
defined by restriction to $S\subset\hg$, are continuous.

The group $G$ acts continuously on ${\rm Hom}\,[\H\G,G]$ in the
following way \cite{AL2}:
%%%%%%%%%%
\be
\label{hoopint6}
{\rm Hom}\,[\H\G,G]\times G\ni(H,g)\mapsto H_g\,:\ H_g(\b)=g^{-1}
H(\b)g,\ \forall\b\in\H\G.
\ee
%%%%%%%%%%
This action corresponds
to the non-trivial part of the action of the group of generalized local
gauge transformations (see below). It is a well established fact that the
quocient space ${\rm Hom}\,[\H\G,G]/G$ is homeomorphic to $\agb$, 
the ``quantum configuration space'' which replaces the 
classical configuration space $\ag$ in the Isham-Ashtekar-Lewandowski
approach to the quantization of theories of 
connections \cite{AI,AL1,AL2,AL3,MM,ALMMT}.

Let us consider now the corresponding action of local gauge transformations
on generalized connections. The group of local gauge transformations
associated with the structure group $G$ is the group $\G$ of all smooth
maps $\gl:\Sig\to G$, acting on smooth connections as follows:
%%%%%%%%%%%%%%%
$$
\A\ni A\mapsto\gl^{-1}A\gl + \gl^{-1}d\gl\,,
$$
%%%%%%%%%%%%%%%
where $d$ denotes the exterior derivative. The corresponding action on
parallel transports $A(\gamma)$ defined by $A\in\A$ and $\gamma\in\E\G$
is given by
%%%%%%%%%%%%%%%%%%%%%%%%%%%%%%%%%%%
\be
\label{hoopint7}
A(\gamma)\mapsto \gl(x_2)^{-1}A(\gamma)\gl(x_1)\, , \ \gl\in\G,
\ee
%%%%%%%%%%%%%%%%%%%%%%%%%%%%%%%%%%%
where $x_1=s(\gamma)$, $x_2=r(\gamma)$. Let us consider the extension
$\bar\G$ of $\G$,
%%%%%%%%%%%%%
\be
\label{hoopint8a}
\bar\G={\rm Map}[\Sigma,G]=G^{\Sigma}\cong\sX_{x\in\Sigma}G_x\,,
\ee
%%%%%%%%%%%%%%%
of all maps $\gl:\Sig\to G$, not necessarily smooth or even continuous. 
This group $\bar\G$ of ``generalized local gauge transformations'' acts
naturally on the space of generalized connections ${\rm Hom}\,[\E\G,G]$,
%%%%%%%%%%%%%%%%
\be
\label{hoopint7a}
{\rm Hom}\,[\E\G,G]\times \bar \G\ni(\bar A,\gl)\mapsto\bar A_{\gl}\in
{\rm Hom}\,[\E\G,G]
\ee
%%%%%%%%%%%%%%%%%
where
%%%%%%%%%%%%%%%%%%%%%%%%%%%%%%%%%%%
\be
\label{hoopint8}
\bar A_{\gl}(\gamma)=
\gl\bigl(r(\gamma)\bigr)^{-1}\bar A(\gamma)\gl\bigl(s(\gamma)\bigr),\ 
\forall\gamma\in\E\G\,,
\ee
%%%%%%%%%%%%%%%%%%%%%%%%%%%%%%%%%%%
generalizing (\ref{hoopint7}). It is natural to consider the quotient of
${\rm Hom}[\E\G,G]$ by the action of $\bar\G$, since ${\rm Hom}[\E\G,G]$
is also made of all the morphisms $\E\G\to G$, without any continuity 
condition. The group $\bar\G$ is compact Hausdorff (with the product 
topology) and its action is continuous \cite{AL2,AL3}. Therefore
${\rm Hom}[\E\G,G]/\bar\G$ is also a compact Hausdorff space.

Let us consider the compact space $\bar\A$ as introduced by Baez, 
e.g.~as the Gelfand spectrum of a commutative unital $C^*$-algebra \cite{B1}. 
According to Gelfand  theory, the original $C^*$-algebra can 
be identified with the
algebra $C(\bar\A)$ of continuous functions in $\bar\A$. The 
group of local gauge transformations
acts on $C(\bar\A)$ and the subspace $C^{\G}(\bar\A)\subset C(\bar\A)$ 
of gauge invariant functions is also a unital commutative $C^*$-algebra,
whose spectrum we will denote by $\bar\A/\bar\G$.

One therefore has four extensions of the classical configuration space 
$\A/\G$, namely $\agb$, $\bar\A/\bar\G$, ${\rm Hom}\,[\H\G,G]/G$ and
${\rm Hom}\,[\E\G,G]/\bar\G$. The first two spaces are tied to the
$C^*$-algebra formalism whereas the last two appear in the context of 
projective methods. As expected, all these spaces are naturally homeomorphic.
Let us consider the following diagram

\medskip
%%%%%%%%%%%%%%%%%%%%%%%%
$$
\begin{array}{ccc}
\agb & \longleftrightarrow & {\rm Hom}\,[\H\G,G]/G \\
& & \\
\updownarrow & & \\
& & \\
\bar\A/\bar\G & \longleftrightarrow & {\rm Hom}\,[\E\G,G]/\bar\G
\end{array}
$$
%%%%%%%%%%%%%%%%%%%%%%%%%%% 
\medskip

\noindent The correspondence between $\agb$ and ${\rm Hom}\,[\H\G,G]/G$ was
established in \cite{MM}. The generalization of this result given in
\cite{AL2} produces a homeomorphism between $\bar\A$ and 
${\rm Hom}\,[\E\G,G]$. It is not difficult to show that this homeomorphism 
is equivariant, leading to a homeomorphism between $\bar\A/\bar\G$ and 
${\rm Hom}\,[\E\G,G]/\bar\G$, as suggested in \cite{AL3}. The
correspondence between $\agb$ and $\bar\A/\bar\G$ follows from results
in \cite{B1}.

In the next subsection we will show directly 
(i.e.~without using the diagram above)
that ${\rm Hom}\,[\E\G,G]/\bar\G$ 
is homeomorphic to ${\rm Hom}[\H\G,G]/G$. The relevance of this 
new proof of a known result lies in 
the clear relation established between ${\rm Hom}\,[\E\G,G]\ (\cong
\bar\A)$ and ${\rm Hom}\,[\H\G,G]/G\ (\cong\agb)$, without having to rely on
the characterization of these spaces as spectra of $C^*$-algebras. 

%%%%%%%%%%%%%%%%%%%%%%%%%%%%%%%%%% Section %%%%%%%%%%%%%%%%%%%%%%%%%%%%%%
\subsection{Equivalence of the projective characterizations of 
$\bar\A/\bar\G$ and $\agb$}
\label{sepc}
%%%%%%%%%%%%%%%%%%%%%%%%%%%%%%%%%%%%%%%%%%%%%%%%%%%%%%%%%%%%%%%%%%%%%%%%%

Since $\H\G\equiv{\rm Hom}\,[x_0,x_0]$ is a subgroup of the groupoid $\E\G$,
a projection  ${\cal P}:{\rm Hom}\,[\E\G,G] \to
{\rm Hom}\,[\H\G,G]$, given by the restriction of elements of 
${\rm Hom}\,[\E\G,G]$ to the group $\H\G$, is naturally defined. 
We will show that this projection
is surjective and equivariant with respect to the actions of $\bar\G$ on 
${\rm Hom}\,[\E\G,G]$ and ${\rm Hom}\,[\H\G,G]$, thus defining a map
${\rm Hom}\,[\E\G,G]/\bar\G\to{\rm Hom}\,[\H\G,G]/G$ 
which is in fact a bijection.
We will also show that the latter map and its inverse are continuous.

We start by  identifying ${\rm Hom}\,[\E\G,G]$ 
with  
${\rm Hom}\,[\H\G,G]\times\bar\G_{x_0}$, where $\bar\G_{x_0}$ is the
subgroup of $\bar\G$ (\ref{hoopint8a}) of the elements $\gl$ such that
$\gl(x_0)={\bf 1}$.
Let us fix a unique edge $e_x\in{\rm Hom}\,[x_0,x]$ for
each $x\in\Sig$, $e_{x_0}$ being the trivial hoop. Let us denote
this set of edges by $\Lambda=\{e_x,\ x\in\Sig\}$. Consider the map
%%%%%%%%%%%%%
\be
\label{novo1}
\Theta_{\Lambda}:{\rm Hom}\,[\E\G,G]\to{\rm Hom}[\H\G,G]\times\bar\G_{x_0}
\ee
%%%%%%%%%%%%% 
where $\bar A\in{\rm Hom}\,[\E\G,G]$ is mapped to $(H,\gl)\in{\rm Hom}\,
[\H\G,G]\times\bar\G_{x_0}$ such that
%%%%%%%%%%%%
\be
\label{novo2}
H(\b)=\bar A(\b),\ \ \forall\b\in\H\G
\ee
%%%%%%%%%%%%
and
%%%%%%%%%%%%%
\be
\label{novo3}
\gl(x)=\bar A\bigl(e_{x}\bigr),\ \ \forall x\in\Sigma\,.
\ee
%%%%%%%%%%%
Consider also the natural action of $\bar\G$ on ${\rm Hom}\,[\H\G,G]\times
\bar\G_{x_0}$ given by
%%%%%%%%%%%%%%%%
\be
\label{hoopint50}
\bigl({\rm Hom}\,[\H\G,G]\times\bar\G_{x_0}\bigr)\times\bar\G\ni\bigl(
(H,\gl),\gl'\bigr)\mapsto (H_{\gl'},\gl_{\gl'})\,,
\ee
%%%%%%%%%%%%%%%
where
%%%%%%%%%%
\be
\label{hoopint51}
H_{\gl'}(\b)=\gl'(x_0)^{-1}H(\b)\gl'(x_0),\ \ \forall\b\in
\H\G
\ee
%%%%%%%%%%%%
and
%%%%%%%%%%%
\be
\label{hoopint52}
\gl_{\gl'}(x)=\gl'(x)^{-1}\gl(x)\gl'(x_0),\ \ \forall x\in\Sigma \,.
\ee
%%%%%%%%%%%%
%%%%%%%%%%%%%%%%%%%%%%%%
\begin{theo}
\label{teonovo1}
For any choice of the set $\Lambda$, the map $\Theta_{\Lambda}$ is a
homeomorphism, equivariant with respect to the action of $\bar\G$.
\end{theo}
%%%%%%%%%%%%%%%%%%%%%%%%%
It is fairly easy to see that $\Theta_{\Lambda}$ is bijective and
equivariant: for a given $\Lambda$, the map $\Theta_{\Lambda}$ is
clearly well defined and its inverse is given by
$(H,\gl)\mapsto\bar A$ where
%%%%%%%%%%%
\be
\label{novo4}
{\bar A}(\gamma)=\gl\bigl(r(\gamma)\bigr)H\Bigl(
e_{r(\gamma)}^{-1}\gamma\,e_{s(\gamma)}
\Bigr)\gl\bigl(s(\gamma)\bigr)^{-1},\ \ \forall\gamma\in\E\G\,.
\ee
%%%%%%%%%
It is also clear that $\Theta_{\Lambda}$ is equivariant with respect to 
the action
of $\bar\G$ on ${\rm Hom}\,[\H\G,G]\times\bar\G_{x_0}$ (\ref{hoopint51}, 
\ref{hoopint52}) and on ${\rm Hom}\,[\E\G,G]$ (\ref{hoopint7a}, 
\ref{hoopint8}). It remains to be
shown that $\Theta_{\Lambda}$ is a homeomorphism. Recall that the
topologies of ${\rm Hom}\,[\H\G,G]$ and ${\rm Hom}\,[\E\G,G]$ are defined by
the projective families $\{\chi_S\}_{S\in\S_{\H}}$ and $\{\A_L\}_{L\in\L}$ 
considered previously.

Given $S\in\S_{\H}$ and $x\in\Sigma$, let $P_S$ and $\pi_x$, respectively, be 
the projections from ${\rm Hom}\,[\H\G,G]\times\bar\G_{x_0}$ to $\chi_S$ and
$G_x$ (the copy of $G$ associated with the point $x$). Recall that the
topology of ${\rm Hom}\,[\H\G,G]\times\bar\G_{x_0}$ 
%%%%%%%%%%%%%%%%
%%is generated by the
%%(inverse images of the) maps $P_S$ and $\pi_x$, meaning that it 
%%%%%%%%%%%%%%%%%%%%%%%%%
is the weakest
topology such that all the maps $P_S$ and $\pi_x$ are continuous. So, $\Theta_
{\Lambda}$ is continuous if and only if the maps $P_S\circ\Theta_{\Lambda}$
and $\pi_x\circ\Theta_{\Lambda}$ are continuous, $\forall S\in\S_{\H}$ and
$\forall x\in\Sigma$. Likewise, $\Theta_{\Lambda}^{-1}$ is continuous if and 
only if all the maps $p_L\circ\Theta_{\Lambda}^{-1}:{\rm Hom}\,[\H\G,G]
\times\bar\G_{x_0}\to\A_L$ are continuous, where the projections 
$p_L:{\rm Hom}\,[\E\G,G]\to\A_L$ are defined in (\ref{eqa1}).

It is straightforward to show that the maps $\pi_x\circ\Theta_{\Lambda}$
are continuous: given $x\in\Sigma$, one just has to consider the
subgroupoid $L=\E\G\bigl\{e_x\bigr\}$ generated by the edge 
$e_x\in\Lambda$ and the homeomorphism (\ref{bara4}) 
$\rho_{e_x}:\A_L\to G$. It is clear that $\pi_x\circ\Theta_{\Lambda}$ 
coincides with $\rho_{e_x}\circ p_L$, being therefore continuous.

On the other hand, to show that $P_S\circ\Theta_{\Lambda}$ and 
$p_L\circ\Theta_{\Lambda}^{-1}$ are continuous one needs to consider 
explicitly the relation between the spaces $\A_L$ and 
$\chi_S$, $L\in\L$, $S\in\S_{\H}$.
%%%%%%%%%%%%%%%%%%%%%%%%%%%%%%%%%%%%%%%%%%%%%%%%%%%%%%%%
\begin{lem}
\label{prophoopint1}
For every $S\in\S_{\H}$ there exists a connected subgroupoid 
$L\in\L$ such that $S$ is a subgroup of $L$. The projection 
%%%%%%%%%%%%%
\be
\label{eqprop1}
p_{S,L}:\A_L\to\chi_S 
\ee
%%%%%%%%%%%%%%
defined by the restriction of elements of $\A_L$ to the subgroup $S$ is 
continuous and satisfies
%%%%%%%%%%%%%
\be
\label{eqprop2}
P_S\circ\Theta_{\Lambda}=p_{S,L}\circ p_L
\ee
%%%%%%%%%%
for every $\Lambda$.
%%%%%%%%%%%%%%%%%%%%
\end{lem}
%%%%%%%%%%%%%%%%%%%%%%%%%%%%%%%%%%%%%%%%%%%%%%%%%%%%%%%%%%%%%%%%%
In order to prove the lemma let us consider a set $\{\b_1,\ldots,\b_n\}$ of
independent hoops generating the group $S$. For each $\b_i$ let us fix a 
piecewise analytic loop $\ell_i$ in the  equivalence class 
$\b_i$ and let $\s_i$ be the corresponding image in \Si. 
We choose a set $\{e_1,\ldots,e_m\}$ of independent edges that decompose
$\cup_{i=1}^n\s_i$, i.e.~$\cup_{i=1}^n\s_i=\cup_{j=1}^m\s(e_j)$,
and denote the connected groupoid
$\E\G\{e_1,\ldots,e_m\}\in\L$ by $L$. 
Since the  hoops $\b_i$ can be obtained as compositions of edges 
$e_j$, $S$ is a subgroup of the  group  ${\rm Hom}_L
\,[x_0,x_0]$ of all arrows of $L$ that start and end at $x_0$.
The generators of $L$ define an homeomorphism (\ref{bara4})
between $\A_L$ and  $G^m$ and the generators of $S$ give us an homeomorphism 
between $\chi_S$ and $G^n$. The same arguments used to prove the continuity
of the maps $p_{L,L'}$ show that the projection
$p_{S,L}:\A_L\to \chi_S$ is continuous (see eq. (\ref{bara3})). 
The relation (\ref{eqprop2}) is obvious.
The  independence with respect to $\Lambda$ follows from the fact that the map
$p_{S,L}$ is independent of  $\Lambda$.~$\Box$

The continuity of the maps $P_S\circ\Theta_{\Lambda}$, $\forall S\in
\S_{\H}$, follows immediately from lemma \ref{prophoopint1}. To show that 
the maps $p_L\circ\Theta_{\Lambda}^{-1}$ are continuous one needs the 
converse of lemma \ref{prophoopint1}. We will use the 
following notation. Given a subgroupoid $\Gamma\subset\E\G$, 
${\rm Obj}\,\Gamma$ denotes the set of objects of $\Gamma$ 
(the set of all points of \Si\ which are
range or source for some arrow in $\Gamma$); ${\rm Hom}_{\,\Gamma}\,[x,y]$
stands for the set of all arrows of $\Gamma$ that start at $x$ and end at
$y$ and $\Pi_{\Gamma}$ denotes the natural projection
from $\bar\G_{x_0}$ to the subgroup $\bar\G_{x_0}(\Gamma)$ of all maps 
${\rm Obj}\,\Gamma\to G$ such that $\gl(x_0)={\bf 1}$. Notice that, as in
theorem \ref{teonovo1}, given a set $\{\gamma_x,\ x\in{\rm Obj}\,\Gamma\}$
of arrows of $\Gamma$, with $\gamma_{x_0}={\bf 1}_{x_0}$, one can define
a bijection between
${\rm Hom}\,[\Gamma,G]$ and ${\rm Hom}\,\bigl[{\rm Hom}_{\,
\Gamma}\,[x_0,x_0],G\bigr]\times\bar\G_{x_0}(\Gamma)$
(in this case we use general arrows instead of edges since some of the sets
${\rm Hom}_{\,\Gamma}\,[x_0,x]$ may not contain any edges).
%%%%%%%%%%%%%%%%%%%%%%%%%%%%%%%%%%%%%%%%%%%%%%%%%%%%%%%%%%%%%%%%%%
\begin{lem}
\label{prophoopint2}
For every $L\in\L$ there exists $S\in\S_{\H}$ and a connected subgroupoid 
$\Gamma\subset\E\G$, with ${\rm Obj}\,\Gamma={\rm Obj}\,L\stackrel{.}{\cup}
\{x_0\}$, such that $L\subset\Gamma$ and ${\rm Hom}_{\,\Gamma}\,
[x_0,x_0]=S$. The natural projection from ${\rm Hom}\,[\Gamma,G]$ to $\A_L$ 
defines a map 
%%%%%%%%%%%
\be
\label{eqprop3}
p_{L,S}:\chi_S\times\bar\G_{x_0}(\Gamma)\to\A_L
\ee
%%%%%%%%%%%%% 
which is continuous and satisfies 
%%%%%%%%%%%%%%
\be
\label{eqprop4}
p_L\circ\Theta_{\Lambda}^{-1}=p_{L,S}\circ(p_S\times\Pi_{\Gamma})
\ee
for an appropriate choice of $\Lambda$.
\end{lem}
%%%%%%%%%%%%%%%%%%%%%%%%%%%%%%%%%%%%%%%%%%%%%%%%%%%%%%%%%%
To prove this lemma let us consider a set $a(L)$ of independent edges
generating the groupoid $L$. If $x_0$ is an object $L$, we take $a(L)$
such that no edges in  $a(L)$ end at $x_0$, which is always possible, reverting
the orientations of some edges if necessary.
Let us consider the subset of ${\rm Obj}\,\Gamma$ of the objects that are
not connected to $x_0$ by an edge in 
$a(L)$. For each such object $x$,
let us add to the set 
$a(L)$ one edge from $x_0$ to  $x$, and denote by
$\bar a(L)$ the set of edges thus obtained.
Of course, one can always choose the new edges such that the set 
$\bar a(L)$ remains independent. 
The image in $\Si$ of the set $\bar a(L)$ is thus a connected graph, and 
$x_0$ is a vertex of this graph. For each object $x$ of $L$, $x\neq x_0$, 
let us choose among the set  $\bar a(L)$ an unique edge from $x_0$ to  $x$,
and call it $e_x$. Let $e_{x_0}$ be the trivial hoop and
$\Lambda(L):=\bigl\{e_x,
\ x\in{\rm Obj}\,L\stackrel{.}{\cup}\{x_0\}\bigr\}$. 
Let $\{e_1,\ldots,e_k\}$ be the subset  of $a(L)$ of the edges that 
do not belong to 
$\Lambda(L)$. With the edges $e_i$ and $e_x$ we construct the hoops
%%%%%%%%%
\be
\label{hoopint55}
\b_i:=e_{r(e_i)}^{-1}e_i e_{s(e_i)},\ \ 
i=1,\ldots,k\,.
\ee
%%%%%%%%%%%
By construction, the set of hoops $\{\b_1,\ldots,\b_k\}$ is independent.
Let $S$ be the  subgroup of $\H\G$ generated  by $\{\b_1,\ldots,\b_k\}$. 
{}From lemma
\ref{lem2}, the set $\Gamma$ of arrows of the form
$e_x\b e_y^{-1}$, with $\b\in S$ and
$x,y\in{\rm Obj}\, L\cup\{x_0\}$, 
is a connected groupoid such that ${\rm Obj}
\,\Gamma={\rm Obj}\,L\stackrel{.}{\cup}\{x_0\}$ 
and ${\rm Hom}_{\,\Gamma}\,
[x_0,x_0]=S$. 
The  groupoid $L$ is a  subgroupoid of $\Gamma$, since all the generators 
of $L$ belong to $\Gamma$, as we show next.
For the edges in $a(L)$ that belong also to $\Lambda(L)$ one has
$e_x=e_x{\bf 1}_{x_0}e_{x_0}^{-1}\in\Gamma$. If, on the other hand,
the edge is of the type $e_i\in\{e_1,\ldots,e_k\}$, then
$e_i=e_{r(e_i)}\b_i e_{s(e_i)}^{-1}\in\Gamma$.
We have therefore proved that there exist $S$ and $\Gamma$ such that 
$L\subset\Gamma$
and ${\rm Hom}_{\,\Gamma}\,[x_0,x_0]=S$.
%%%%%%%%%%%%%%%%%%%%%%%%%%%%%%%%%%%%%%%%%%%%%%%%%%%%%%%
%%%%%%%%%%%%%%%%%%%%%%%%%%%%%%%%%%%%%%%%%%%%
For the remaining of the proof, let 
$p_{L,\Gamma}:{\rm Hom}\,[\Gamma,G]\to\A_L$ be the projection defined by  
restriction to  $L$ and let
%%%%%%%%%%%
\be
\label{hoopint57}
\Theta_{\Lambda(L)}(\Gamma):{\rm Hom}\,[\Gamma,G]\to\chi_S\times\bar
\G_{x_0}(\Gamma)
\ee
%%%%%%%%%%%%%
be the bijection associated to the set $\Lambda(L)$. We introduce also the
notation 
%%%%%%%%%%%%%
\be
\label{hoopint58}
p_{L,S}:=p_{L,\Gamma}\circ\Theta_{\Lambda(L)}^{-1}(\Gamma):\chi_S\times\bar
\G_{x_0}(\Gamma)\to\A_L\,.
\ee
%%%%%%%%%%%%%%
Since $\chi_S\times\bar\G_{x_0}(\Gamma)$ and $\A_L$ 
can be identified with powers of $G$, we conclude, as in the proof of lemma 
\ref{prophoopint1}, that 
$p_{L,S}$ is continuous. Finally, to prove 
(\ref{eqprop4}) one just has to consider a set of edges $\Lambda$ that
contains $\Lambda(L)$.~$\Box$

Given that the projections $p_S:{\rm Hom}\,[\H\G,G]\to\chi_S$ and 
$\Pi_{\Gamma}:\bar\G_{x_0}\to\bar\G_{x_0}(\Gamma)$ are continuous, lemma
\ref{prophoopint2} shows that for every fixed $L\in\L$ there exists a  
$\Lambda$ such that $p_L\circ\Theta_{\Lambda}^{-1}$ is continuous, 
which 
still does not prove that all the maps $p_L\circ\Theta_{\Lambda}^{-1}$,
$L\in\L$, are continuous for a given $\Lambda$. We have however:
%%%%%%%%%%%%%%%%%%%%%%%%%%%%%%%%%%%%%%%%%%%%%
\begin{lem}
\label{lemhoopint3}
The map $\Theta_{\Lambda}\circ\Theta_{\Lambda'}^{-1}$ is a homeomorphism
for any $\Lambda$ and $\Lambda'$.
\end{lem}
%%%%%%%%%%%%%%%%%%%%%%%%%%%%%%%%%%%%%%%%%%%%%%
To prove this result notice that the map
%%%%%%%%%%%%%
\be
\label{hoopint59}
\Theta_{\Lambda}\circ\Theta_{\Lambda'}^{-1}:{\rm Hom}\,[\H\G,G]\times\bar
\G_{x_0}\to{\rm Hom}\,[\H\G,G]\times\bar\G_{x_0}
\ee
%%%%%%%%%%%%%%
is given by
%%%%%%%%%%
\be
\label{hoopint60}
(H,\gl)\mapsto(H',\gl')
\ee
%%%%%%%%%%%%
such that
%%%%%%%%%%%%%%
\be
\label{hoopint61}
H'=H,\ \ \gl'(x)=\gl(x)H\bigl(e_x^{-1}e'_x\bigr),\ \ \forall
x\in\Sigma\,,
\ee
%%%%%%%%%%%%
where $e_x\in\Lambda$ and $e'_x\in\Lambda'$. It is then
sufficient to show that the maps $\pi_x\circ\Theta_{\Lambda}\circ\Theta_
{\Lambda'}^{-1}$ are continuous, $\forall x\in\Sigma$, since
$P_S\circ\Theta_{\Lambda}\circ\Theta_{\Lambda'}^{-1}=P_S$, $\forall S\in\S_
{\H}$. But $\pi_x\circ\Theta_{\Lambda}\circ\Theta_{\Lambda'}^{-1}$ can be  
obtained as composition of the maps
%%%%%%%%%%
\be
\label{hoopint62}
{\rm Hom}\,[\H\G,G]\times\bar\G_{x_0}\ni(H,\gl)\mapsto\Bigl(H\bigl(
e_x^{-1}e'_x\bigr),\gl(x)\Bigr)\in G\times G
\ee
%%%%%%%%%%%
and
%%%%%%%%%%%%%
\be
\label{hoopint63}
G\times G\ni(g_1,g_2)\mapsto g_2g_1\in G\,,
\ee
%%%%%%%%%%%%
which are clearly continuous.~$\Box$

{}From lemma \ref{lemhoopint3} follows immediately that
%%%%%%%%%%%%%%%%%%%%%%%%%%%%%%%%%%%%%%%%%%%%%%%
\begin{cor}
\label{corhoopint2}
The continuity of $p_L\circ\Theta_{\Lambda}^{-1}$ is equivalent to the
continuity of $p_L\circ\Theta_{\Lambda'}^{-1}$, for any other values of 
$\Lambda'$.
\end{cor}
%%%%%%%%%%%%%%%%%%%%%%%%%%%%%%%%%%%%%%%%%%%%%%%
This corollary, together with lemma \ref{prophoopint2}, shows that, for a 
given $\Lambda$, all the maps $p_L\circ\Theta_{\Lambda}^{-1}$, $L\in\L$, 
are continuous, which concludes the proof of theorem \ref{teonovo1}.~$\Box$

\smallskip

The identification of ${\rm Hom}\,[\E\G,G]/\bar\G$ with ${\rm Hom}\,[\H\G,G]/G$
now follows easily. Consider a fixed $\Lambda$. Since $\Theta_{\Lambda}$
is a homeomorphism equivariant with respect to the continuous 
action of $\bar\G$, we conclude that
${\rm Hom}\,[\E\G,G]/\bar\G$ is homeomorphic to
$\bigl({\rm Hom}\,[\H\G,G]\times\bar\G_{x_0}\bigr)/\bar\G$. On the other
hand it is clear that
%%%%%%%%%%%%%%
\begin{eqnarray}
\bigl({\rm Hom}\,[\H\G,G]\times\bar\G_{x_0}\bigr)/\bar\G & = &
\bigl({\rm Hom}\,[\H\G,G]/G\bigr)\times\bigl(\bar\G_{x_0}/\bar\G_{x_0}\bigr)
\cong \nonumber \\
\label{numsegunda}
&\cong & {\rm Hom}\,[\H\G,G]/G\,.
\end{eqnarray}
%%%%%%%%%%%%%%
Thus, as a corollary of theorem \ref{teonovo1} one gets that
%%%%%%%%%%%%%%%%%%%%%%%%%%%%%%%%%%%%%%%%%%%%%%%%%%%%%%%%%%%
\begin{theo}
\label{corhoopint1}
The spaces ${\rm Hom}\,[\E\G,G]/\bar\G$ and ${\rm Hom}\,[\H\G,G]/G$ are
homeomorphic.
\end{theo}
%%%%%%%%%%%%%%%%%%%%%%%%%%%%%%%%%%%%%%%%%%%%%%%%%%%%%%%%%%%
It is also interesting to note that the identification ${\rm Hom}\,[\H\G,G]
\times \bar\G_{x_0}\cong{\rm Hom}\,[\E\G,G]$, through the choice of a set
of edges $\Lambda=\bigl\{e_x,\ x\in\Sig\bigr\}$ as above, provides a
(almost) global gauge-fixing, meaning that 
there are sections $\eta : {\rm Hom}\,[\H\G,G]\to {\rm Hom}\,[\E\G,G]$
such that ${\cal P}\circ \eta = {\bf id}$, where 
${\cal P}:{\rm Hom}\,[\E\G,G]\to {\rm Hom}\,[\H\G,G]$
is the canonical projection.
${\rm Hom}\,[\H\G,G]$ can therefore be
identified with a subset of ${\rm Hom}\,[\E\G,G]$. In fact, since the edges 
$e_x$ in the set $\Lambda$ are algebraically independent, the
space ${\rm Hom}\,[\H\G,G]$ can be seen as a subset of ${\rm Hom}\,[\E\G,G]$
of all generalized connections with given preassigned values on the set
$\Lambda$. Choosing, for instance, the identity of $G$ for all $e_x$,
one then has the identification
%%%%%%%%%%%%%%%%
\begin{equation}
\label{gfix}
{\rm Hom}\,[\H\G,G]\cong\Bigl\{
\bar A\in{\rm Hom}\,[\E\G,G]\ \ |\ \ \bar A\bigl(e_x\bigr)={\bf 1},
\ \ \forall x\in\Sig\Bigr\}\,.
\end{equation}
%%%%%%%%%%%%%%%%
There remains, of course, the non-trivial
action of gauge
transformations at the base point $x_0$.
A study of the action of
the full gauge group  $\bar\G$
was
recently done by Fleischhack, leading to
stratification results in the
context of generalized connections \cite{F1,F3}
(see also \cite{V}). A 
detailed account on the existence
of Gribov ambiguities when the full
gauge-invariant space ${\rm Hom}\,[\H\G,G]/G\cong{\rm Hom}\,[\E\G,G]/\bar\G$
is considered is given in \cite{F4}.

%%%%%%%%%%%%%%%  Acknowledgments %%%%%%%%%%%%%%%%%%%%%%

\section*{Acknowledgements}
\noindent I would like to thank Jos\'e Mour\~ao, Paulo S\'a and 
Thomas Thiemann, for
encouragement and helpful discussions. This work was supported in part by 
PRAXIS 2/2.1/FIS/286/94,
CERN/P/FIS/15196/1999 and CENTRA/UAlg.
%%%%%%%%%%%%%%%%%%%%%%%%%%%%%%%%

%%%%%%%%%%%%%% Bibliography %%%%%%%%%%%%%%%%%%%%%%%

%%%%%%%%%%%%%%%%%%%%%%%%%%%%%%%%%%

%%%%%%%%%%%%%%%%%%%%%%%%%%%%%%%%%

\end{document}